\documentclass[12pt]{iopart}
\usepackage{iopams}

\usepackage{graphicx}
\usepackage{pstricks}
\usepackage{pst-node}

\begin{document}

\title{Perturbation theory via Feynman diagrams in classical mechanics}

\author{R. Penco, D. Mauro$^{\dagger}$}
\address{%
Department of Theoretical Physics, 
	     University of Trieste, \\
	     Strada Costiera 11, Miramare-Grignano, 34014 Trieste, Italy\\
	     $^{\dagger}$and INFN, Trieste, Italy} 
\ead{penco@ts.infn.it, mauro@ts.infn.it}%

\begin{abstract}
In this paper we show how Feynman diagrams, which are used as a tool to implement perturbation theory in quantum field theory, can be very useful also in classical mechanics, provided we introduce also at the classical level concepts like path integrals and generating functionals.  
\end{abstract}

\maketitle

\section{Introduction}

In physics different theories are often formulated in different languages. For example, quantum mechanics is formulated via states and operators in a Hilbert space endowed with a commutator structure.
Classical mechanics for a system with $n$ degrees of freedom is instead formulated in a phase space labeled by $n$ coordinates $q^j$ and $n$ canonical momenta $p^j$, which we will collect together in a unique $\varphi^a\equiv(q^j,p^j)$, with the index $a$ running from 1 to $2n$. This phase space is endowed with a Poisson brackets (pb) structure, which can be written as: $\{\varphi^a,\varphi^b\}_{\textrm{\scriptsize{pb}}}=\omega^{ab}$, where the symplectic matrix $\omega^{ab}$ is the following $2n\times 2n$ antisymmetric matrix: 
$\omega=\pmatrix{0 & 1 \cr -1 & 0}$. 

In most textbooks a theory is usually developed in only one way and the alternative formulations of the same theory are often not mentioned at all. All this can lead students to the misconception that there is a very strict link between a theory and its mathematical formulation. Of course, there is often a particular language which is more suitable than others to express a physical theory, but it is important to realize that such a language is often not the only possible one. In this paper we want to stress the distinction between the physical content of a theory and the language which is commonly used to formulate the theory itself. In fact, we will provide a simple example to show explicitly how the {\it same language} (the path integrals) can be used to formulate {\it different theories}, like classical and quantum mechanics.

As a consequence, we shall also show that the well-known {\it Feynman diagrams}, used mostly in quantum field theory to develop perturbation theory, can also be used in classical mechanics to provide a perturbative solution to the classical equations of motion. This paper can then be useful as an introduction to the main ideas and concepts of Feynman diagrammatics, which are studied in quantum field theory courses.

\section{A brief introduction to a path integral approach for classical mechanics}

As we said in the Introduction, the main purpose of this paper is to implement perturbation theory in classical mechanics via the same tools that are used in quantum field theory. To reach this aim we have first of all to introduce also in classical mechanics concepts like path integrals and generating functionals. 

It is well-known that path integrals were first introduced in quantum mechanics by R.P. Feynman \cite{hibbs}. He discovered that the transition amplitude of going from a certain point $q_i$ at the initial time $0$ to another point $q_f$ at the final time $T$ can be written as a sum over the paths joining these two points on an interval of time $T$. In particular, it is possible to show that all the paths in phase space give a contribution to such a transition amplitude according to the following formula:
\begin{equation}
\displaystyle K(q_f,q_i;T)=\int_{q_i}^{q_f} {\mathcal D}^{\prime\prime}q{\mathcal D}p\, \exp \left[ i\int_{0}^{T}\, \textrm{d}t \, L\right], \label{qpi}
\end{equation}
where the double prime over ${\mathcal D}$ indicates that the functional integration is over functions $q(t)$ with fixed end points, i.e. $q(0)=q_i$ and $q(T)=q_f$, and $L$ is the following Lagrangian:
\begin{equation}
\displaystyle L=p\dot{q}-H, \qquad H=\frac{p^2}{2}+V(q). \label{1bis}
\end{equation} 

Let us now pass to the path integral formulation of classical mechanics developed at the end of the 1980s \cite{ennio}. The starting point of this formulation is the following: which is the probability density $P(\varphi_f,\varphi_i;T)$ of going from a certain point of the phase space $\varphi_i$ at time $0$ to another point $\varphi_f$ at time $T$? Of course, we know that the answer is very different from the quantum case. In fact, in classical mechanics, once we fix the initial conditions $\varphi(0)=\varphi_i$, there is only one path  in phase space which solves the equations of motion, let us indicate it with $\varphi^a_{\textrm{\scriptsize{cl}}}(t;\varphi_i)$, where ``cl" stands for ``classical path". As a consequence, the probability density $P(\varphi_f, \varphi_i; T)$ is different than zero only when the point $\varphi_f$ lies on the classical path $\varphi^a_{\textrm{\scriptsize{cl}}}(T;\varphi_i)$:
\begin{equation}
\displaystyle P(\varphi_f,\varphi_i;T)=\delta[\varphi_f^a-\varphi_{\textrm{\scriptsize{cl}}}^a(T;\varphi_i)].  \label{2bis}
\end{equation}
Now we want to rewrite the previous result as a path integral over all the possible trajectories in phase space $\varphi(t)$, where the only path which contributes is the classical one. This can be realized via a suitable functional Dirac delta, i.e.: 
\begin{equation}
\displaystyle P(\varphi_f,\varphi_i;T)=\int_{\varphi_i}^{\varphi_f} {\mathcal D}^{\prime\prime}\varphi\,{\delta}[\varphi^{a}-\varphi^{a}_{\textrm{\scriptsize{cl}}}(t;\varphi_i)]. \label{cpi2}
\end{equation} 
This is a first expression for the path integral for classical mechanics \cite{ennio}. Let us  now remember the following property of the Dirac deltas: if a function $f(\varphi)$ has a zero in $\varphi_0$, then the following equation holds: 
$\displaystyle \delta(\varphi-\varphi_0)=\delta[f(\varphi)] \cdot |f^{\prime}(\varphi_0)|.$
In our case the function $f(\varphi)$ is given by the LHS of the classical Hamilton equations of motion $\dot{\varphi}^a-\omega^{ab}\partial_bH=0$. As a consequence, the zero $\varphi_0$ of $f(\varphi)$ becomes the path $\varphi^a_{\textrm{\scriptsize{cl}}}(t;\varphi_i)$ which solves the classical equations of motion. The modulus of $f^{\prime}$ becomes instead a functional determinant and the weight of the classical path integral (\ref{cpi2}) can then be rewritten as follows:
\begin{equation}
\delta[\varphi^a-\varphi^a_{\textrm{\scriptsize{cl}}}(t;\varphi_i)]=\delta(\dot{\varphi}^a-\omega^{ab}\partial_bH)
\textrm{det}(\partial_t\delta^a_b-\omega^{ac}\partial_c\partial_bH). 
\label{zeros}
\end{equation}
It can be proved \cite{ennio} that the functional determinant in the previous formula is independent of the phase space variables $\varphi$, so it can be omitted at this level. Using the Fourier transform, the Dirac delta of the equations of motion can be rewritten as a path integral over the auxiliary variables $\lambda_a\equiv (\lambda_q,\lambda_p)$:
\begin{displaymath}
\displaystyle \delta(\dot{\varphi}^a-\omega^{ab}\partial_bH)=\int {\mathcal D}\lambda
\, e^{i\int_0^T \textrm{\scriptsize{d}}t\, \lambda_a(\dot{\varphi}^a-\omega^{ab}\partial_bH)}.
\end{displaymath}
So the path integral for classical mechanics \cite{ennio} becomes a functional integral over an extended phase space $(\varphi, \lambda)$:
\begin{equation}
\displaystyle
P(\varphi_f,\varphi_i; T)=\int_{\varphi_i}^{\varphi_f}{\mathcal D}^{\prime\prime}\varphi{\mathcal D}\lambda\,
\textrm{exp}\left[i\int_{0}^{T} {\textrm{{d}}}t\,{\mathcal L} \right], \label{cpi}
\end{equation}
where the Lagrangian and the Hamiltonian associated with each path are given by:
\begin{equation}
{\mathcal L}=\lambda_a\dot{\varphi}^a-{\mathcal H}, \qquad 
{\mathcal H}=\lambda_a\omega^{ab}\partial_bH.  \label{4bis}
\end{equation}
In this way we have rewritten the weight of the path integral for classical mechanics in an exponential form, just like in the quantum case. The main differences between the two path integrals for quantum and classical mechanics (\ref{qpi}) and (\ref{cpi}) are in the space of functions which are integrated over and in the arguments of the exponential weight, which are defined in (\ref{1bis}) and (\ref{4bis}) respectively.
Nevertheless, since we have classical mechanics formulated in the same language of quantum mechanics, it is reasonable to expect that, even in classical mechanics, we can develop perturbation theory via Feynman diagrams. 

\section{The classical generating functional}

In Eq. (\ref{cpi}) we have found a path integral for the classical transition probabilities. What we want to do now is to define the analog of the generating functional \cite{book} for classical mechanics. Let us introduce two currents, $J$ and $\Lambda$, associated with $q$ and $\lambda_p$ respectively, and define: 
\begin{equation}
\displaystyle Z[J,\Lambda]\equiv \int
\textrm{d}\varphi_f \int_{\varphi_i}^{\varphi_f}{\mathcal D}^{\prime\prime}\varphi{\mathcal D}\lambda \,
e^{i\int_{0}^{T}\textrm{\scriptsize{d}}t \, ({\mathcal L}+Jq+\Lambda\lambda_p)}.  \label{zeta}
\end{equation}
When $J=\Lambda=0$ the generating functional $Z[J,\Lambda]$ 
is normalized to one. In fact, using (\ref{cpi}) and (\ref{2bis}):
\begin{eqnarray*}
\displaystyle Z[J,\Lambda]\Big|_{J,\Lambda=0}&=&\int \textrm{{d}}\varphi_f P(\varphi_f,\varphi_i;T)
\nonumber \\
\displaystyle &=& \int \textrm{{d}}\varphi_f \, \delta[\varphi^a_f- \varphi^{a}_{\textrm{\scriptsize{cl}}}(T;\varphi_i)]=1.
\end{eqnarray*}
The generating functional $Z[J,\Lambda]$ can be used to formally write down the solutions of the equation of motion. In fact, as we will prove in Appendix A, the functional derivative of $Z[J,\Lambda]$ w.r.t. $J(t)$ reproduces exactly the solution of the equations of motion associated with the initial conditions 
$\varphi(0)=\varphi_i$: 
\begin{equation}
\displaystyle -i\frac{\delta Z[J,\Lambda]}{\delta J(t)} \Biggl|_{J,\Lambda=0}=q_{\textrm{\scriptsize{cl}}}(t;\varphi_i), \quad 0<t<T.   \label{firstappendix}
\end{equation}
Of course, for a generic potential it is impossible to calculate exactly  the path integral (\ref{zeta}) and consequently its derivative (\ref{firstappendix}) to get the exact expression for $q_{\textrm{\scriptsize{cl}}}(t;\varphi_i)$. Nevertheless, as we will prove in Appendix B, we can derive the explicit form of the generating functional $Z_0[J,\Lambda]$ for a harmonic oscillator described by the Hamiltonian:
\begin{displaymath}
\displaystyle H_0=\frac{1}{2}p^2+\frac{1}{2}\omega^2q^2. 
\end{displaymath}
The result is:
\begin{equation}
\fl \qquad \qquad \displaystyle Z_0[J,\Lambda]=\exp \left[ i \int_{0}^{T} \textrm{{d}}t \, J(t)q_{0}(t;\varphi_i)\medskip  \label{zedzero}  -
i \int_0^T \textrm{d}t \, \textrm{d}t^{\prime}\, J(t)G_{\textrm{\scriptsize{R}}}(t-t^{\prime})\Lambda(t^{\prime}) \right], \nonumber
\end{equation}
where $q_{0}(t;\varphi_i)$ is the solution of the harmonic oscillator equations of motion with initial conditions $\varphi_i\equiv (q_i,p_i)$:
\begin{equation}
\displaystyle q_{0}(t;\varphi_i)=q_i\cos \omega t +\frac{p_i}{\omega}\sin \omega t \label{harm}
\end{equation}
and $G_{\textrm{\scriptsize{R}}}$ is the following retarded Green function (see Appendix B for more details): 
\begin{equation}
G_{\textrm{\scriptsize{R}}}(t-t^{\prime})\equiv \Theta(t-t^{\prime})\frac{\sin \omega(t-t^{\prime})}{\omega}.  \label{green}
\end{equation}
As a consequence, from Eq. (\ref{zedzero}) we have that 
\begin{equation}
\displaystyle -i\frac{\delta Z_0[J,\Lambda]}{\delta J(t)}\Biggl|_{J,\Lambda=0}=q_{0}(t;\varphi_i), \qquad 0<t<T, \label{minni}
\end{equation}
which confirms the result (\ref{firstappendix}) in the case of a harmonic oscillator. 
In a similar way, we can get the retarded propagator (\ref{green}) by acting on the generating functional with two functional derivatives as follows:
\begin{equation}
\displaystyle (-i)^2\frac{\delta^2 Z_0[J,\Lambda]}{\delta J(t)\delta \Lambda(t^{\prime})}\Biggl|_{J,\Lambda=0}=iG_{\textrm{\scriptsize{R}}}(t-t^{\prime}). \label{topolino}
\end{equation}

The generating functional (\ref{zedzero}) can be considered as a starting point to implement perturbation theory for classical mechanics. In particular, we will call {\it propagators} the quantities in Eqs. (\ref{minni})-(\ref{topolino}), which are coupled with the external currents in the generating functional. These propagators can be graphically represented according to the following Feynman rules:
\begin{center}
\begin{picture}(120,40)
\put(0,9){\line(1,0){30}}  
\put(34,9){\line(1,0){8}}  
\put(46,9){\line(1,0){8}} 
\put(58,9){\line(1,0){8}} 
\put(80,7){$=\;\, iG_\textrm{\scriptsize{R}}(t-t^{\prime})$.}
\put(0,9){\circle*{3.5}}
\put(68,9){\circle*{3.5}}
\put(20,33){\line(1,0){34}}
\put(20,33){\circle*{3.5}}
\put(80,31){$=\;\, q_{0}(t;\varphi_i)$}
\end{picture}
\end{center}
\vspace{-0.15cm}

Next, let us consider a system described by a Hamiltonian $H$ which can be split into the sum of the Hamiltonian $H_0$ of the harmonic oscillator and a term of perturbation $gV(q)$:
\begin{displaymath}
\displaystyle H=H_0+gV(q).
\end{displaymath}
Then the ${\mathcal H}$ of Eq. (\ref{4bis}) can be split as:
\begin{displaymath}
\displaystyle {\mathcal H} = \lambda_a\omega^{ab}\partial_bH_0-g\lambda_pV^{\prime}(q) \equiv {\mathcal H}_0+g{\mathcal V}(q,\lambda_p),
\end{displaymath}
where in the first step we have used the fact that the perturbation $V$ is independent of $p$. The generating functional associated with the system described by the new Hamiltonian $H$ can then be written in terms of the free generating functional (\ref{zedzero}) as follows:
\begin{eqnarray}
\fl \qquad \displaystyle Z= \int \textrm{d}\varphi_f \int_{\varphi_i}^{\varphi_f}{\cal D}^{\prime\prime}\varphi {\cal D}\lambda
\exp\left[ i\int_{0}^{T} \textrm{d}t\, (\lambda_a\dot{\varphi}^a -{\mathcal H}_0
-g{\mathcal V}(q,\lambda_p)+Jq+\Lambda\lambda_p)\right]  \label{gibbs} \\
\fl \qquad \quad = \displaystyle \exp \left[-ig\int_{0}^{T}\textrm{d}t \, {\mathcal V}\left( -i\frac{\delta}{\delta J(t)}, -i\frac{\delta}{\delta \Lambda(t)}\right)\right] Z_0. \nonumber
\end{eqnarray}
In the last line of Eq. (\ref{gibbs}) we have replaced the arguments of ${\mathcal V}$, i.e. $q$ and $\lambda_p$, with the functional derivatives w.r.t. the associated currents $J$ and $\Lambda$ respectively. 

According to Eq. (\ref{gibbs}), the generating functional $Z$ can be approximately calculated by expanding the differential operator acting on $Z_0$ up to the desired order in $g$. As we will see in the next section, this will lead to another Feynman rule which crucially depends on the details of the perturbation. 

\section{An example: the anharmonic oscillator}

We will now consider an anharmonic oscillator. In this case the perturbation is $\displaystyle V(q)=\frac{q^4}{24}$, which implies: 
$\displaystyle {\mathcal V}(q,\lambda_p)=-\frac{1}{6}\lambda_pq^3.$ The effect of this perturbation can be taken into account by introducing a new graphical element known as {\it vertex}. 
In particular, we will associate every factor in the perturbation ${\mathcal V}$ with an external leg: a dotted one, to represent $\lambda_p$, and three full ones, to represent each factor $q$.
In conclusion, the Feynman rule for the vertex is the following one:
\vspace{-0.4cm}
\begin{center}
\begin{figure}[!h]
\hspace{4.8cm}
\pspicture(0.25,0.25)(1.75,1.75)
\psset{linewidth=0.5pt}
\psline[]{-}(0.25,1.75)(1.75,0.25)
\psline[]{-}(0.25,0.25)(1,1)
\psline[]{-}(1,1)(1.15,1.15)
\psline[]{-}(1.3,1.3)(1.45,1.45)
\psline[]{-}(1.6,1.6)(1.75,1.75)
\pscircle*(1,1){0,06}
\endpspicture 
\end{figure}  \vspace{-1.85cm} 
$\displaystyle =i\, \frac{g}{6} \int_{0}^{T} \textrm{d}t.$
\end{center}
\vspace{0.4cm}

Now, as a first example, let us calculate, up to the first order in $g$, the solution of the classical equation of motion for an anharmonic oscillator.
First of all, let us indicate with a circle
the generating functional $Z[J,\Lambda]$, which has the following diagrammatic expression:
\vspace{-0.4cm}
\begin{center}
\begin{picture}(200,60)
\put(70,30){\circle{30}} 
\put(2,28){$Z[J,\Lambda]=$}
\put(124,30){\line(1,0){5}}
\put(133,30){\line(1,0){5}}
\put(142,30){\line(1,0){5}}
\put(148,30){\circle*{3}}
\put(148,30){\line(1,1){15}}
\put(148,30){\line(1,0){20}}
\put(148,30){\line(1,-1){15}}
\put(173,28){$+O(g^2).$}
\put(90,28){$= 1\, +$}
\end{picture}
\end{center}
\vspace{-0.4cm}
The functional derivative w.r.t. $J(t)$ can be indicated as:
\begin{center}
\begin{picture}(100,60)
\put(95,30){\circle{30}} 
\put(49,30){\line(1,0){30}}
\put(49,30){\circle*{3.5}}
\put(47,16){$t$}
\put(-50,28){$-i\displaystyle \frac{\delta Z[J,\Lambda]}{\delta J(t)}\Biggl|_{J,\Lambda=0}=$}
\put(120,28){.}
\end{picture}
\end{center}
\vspace{-0.4cm}
Using Eq. (\ref{firstappendix}) we have that the solution of the classical equations of motion for an anharmonic oscillator can be represented via the following Feynman diagram:
\vspace{-0.4cm}
\begin{center}
\begin{picture}(175,60)
\put(93,30){\circle{30}} 
\put(47,30){\line(1,0){30}}
\put(47,30){\circle*{3.5}}
\put(45,16){$t$}
\put(2,28){$q_{\textrm{\scriptsize{cl}}}(t)=$}
\put(2,-13){$\phantom{q_{\textrm{\scriptsize{cl}}}(t)}=$}
\put(47,-11){\line(1,0){30}}
\put(47,-11){\circle*{3}} 
\put(45,-22){$t$}
\put(82,-13){$+$}
\put(98,-11){\line(1,0){15}}
\put(98,-11){\circle*{3}}
\put(96,-22){$t$}
\put(116,-11){\line(1,0){4}}
\put(123,-11){\line(1,0){4}}
\put(130,-11){\line(1,0){4}}
\put(136,-11){\circle*{3}}
\put(136,-11){\line(1,1){10}}
\put(136,-11){\line(1,0){15}}
\put(136,-11){\line(1,-1){10}}
\put(153,-13){$\; + \; O(g^2)$,}
\end{picture}
\bigskip
\bigskip
\bigskip
\end{center}
\vspace{-0.4cm}
which, using our Feynman rules, can be translated in the following equation:
\begin{displaymath}
\displaystyle q_{\textrm{\scriptsize{cl}}}(t)=q_0(t)-\frac{g}{6}\int_0^{T}\textrm{d}t^{\prime}\,
G_{\textrm{\scriptsize{R}}}(t-t^{\prime})q_0^3(t^{\prime})+O(g^2).
\end{displaymath}
Let us solve explicitly the previous equation, considering for simplicity the particular initial condition $p_i=0$. Then, using (\ref{harm})-(\ref{green}), we have the following solution for the classical equation of motion of the anharmonic oscillator:
\begin{equation}
\displaystyle q_{\textrm{cl}}(t)=q_{0}(t)-\frac{gq_i^3}{48\omega^2} \biggl[ 3\omega t
\sin \omega t +\frac{1}{4} \left( \cos \omega t-\cos 3\omega t\right)\biggr]+O(g^2). \label{equivalence}
\end{equation}
This is the same result that can be obtained using the standard method of secular perturbation theory, see for example \cite{jose'}.

\section{Equivalence with the secular perturbation theory} 

The equivalence of the result (\ref{equivalence}) with the one that can be obtained via the secular perturbation theory is not an accident. In fact, it is quite easy to show that our Feynman diagrammatics is just a stenographic expression for the secular approach to perturbation theory in classical mechanics. In fact, let us consider the classical equation of motion for the anharmonic oscillator:
\begin{equation}
\displaystyle \ddot{q}(t) +\omega^2q(t)=-\frac{g}{6}q^3(t). \label{eqmotion}
\end{equation}
The secular approach to perturbation theory in classical mechanics \cite{jose'} consists, first of all, in expanding $q(t)$ in series of $g$:
\begin{displaymath}
\displaystyle q(t)=q_0(t)+gq_1(t)+g^2q_2(t)+\cdots.
\end{displaymath}
Replacing the previous expansion in the equation of motion (\ref{eqmotion}) we get an infinite set of coupled equations for the components $q_k(t)$:
\begin{eqnarray}
\displaystyle \ddot{q}_0(t)+\omega^2q_0(t) &=& 0, \nonumber \medskip \\
\displaystyle \ddot{q}_1(t)+\omega^2q_1(t) &=& -\frac{1}{6}q_0^3(t),  \label{eqcomp} \medskip \\
\displaystyle \ddot{q}_2(t)+\omega^2q_2(t) &=& -\frac{1}{2}q_0^2(t)q_1(t). \nonumber
\end{eqnarray}
Then the boundary conditions $q(0)=q_i$ and $\dot{q}(0)=p_i$ can be turned into the following set of conditions:
\begin{equation}
\begin{array}{l}
\displaystyle q_0(0)=q_i, \qquad \dot{q}_0(0)=p_i, \medskip \\ 
q_k(0)=\dot{q}_k(0)=0 \qquad \forall k\ge 1, \label{bcond}
\end{array}
\end{equation}
which can be used to solve the equations of motion (\ref{eqcomp}). The first equation is the unperturbed equation, which can be solved to find the $q_{0}$ of Eq. (\ref{harm}). This expression can be inserted into the second equation of (\ref{eqcomp}). Imposing the boundary conditions (\ref{bcond}), one then finds the solution for $gq_1(t)$:
\begin{equation}
\displaystyle gq_1(t)=-\frac{g}{6} \int_{0}^{T}\textrm{d}t^{\prime}\, G_{\textrm{\scriptsize{R}}}(t-t^{\prime})q_0^3(t^{\prime}), \label{giuno}
\end{equation}
which is just the contribution given by the following Feynman diagram:
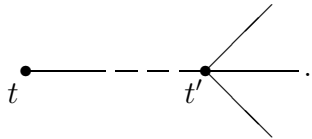
\begin{center}
\begin{picture}(100,60)
\put(-7,21){$t$}
\put(0,33){\line(1,0){30}}  
\put(34,33){\line(1,0){8}}  
\put(46,33){\line(1,0){8}} 
\put(58,33){\line(1,0){8}} 
\put(60,21){$t^{\prime}$}
\put(68,33){\line(1,1){25}}
\put(68,33){\line(1,0){35}}
\put(68,33){\line(1,-1){25}}
\put(0,33){\circle*{3.5}}
\put(68,33){\circle*{3.5}} \put(105,31){.}
\end{picture}
\end{center}
\vspace{-0.3cm}
This procedure can be iterated. For example, using (\ref{giuno}) in the third of Eqs. (\ref{eqcomp}), we get:
\begin{eqnarray*}
g^2q_2(t)&=& -\frac{g^2}{2}\int_{0}^{T}\textrm{d}t^{\prime}\, G_{\textrm{\scriptsize{R}}}(t-t^{\prime})q_0^2(t^{\prime})q_1(t^{\prime}) \nonumber\\
&=& \frac{g^2}{12}\int_{0}^{T}\textrm{d}t^{\prime}\, G_{\textrm{\scriptsize{R}}}(t-t^{\prime})q_0^2(t^{\prime})\cdot \int_{0}^{T}
\textrm{d}t^{\prime\prime}\, G_{\textrm{\scriptsize{R}}}(t^{\prime}-t^{\prime\prime})q_0^3(t^{\prime\prime})
\end{eqnarray*}
which corresponds to the following Feynman diagram multiplied by 3:
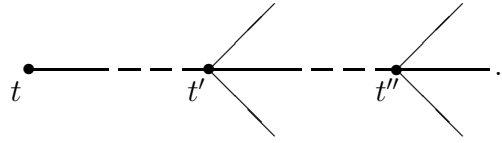
\begin{center}
\hspace{-2cm}
\begin{picture}(100,60)
\put(0,33){\line(1,0){30}}  
\put(34,33){\line(1,0){8}}  
\put(46,33){\line(1,0){8}} 
\put(58,33){\line(1,0){8}} 
\put(68,33){\line(1,1){25}}
\put(68,33){\line(1,0){35}}
\put(68,33){\line(1,-1){25}}
\put(0,33){\circle*{3.5}}
\put(-7,21){$t$}
\put(68,33){\circle*{3.5}}
\put(60,21){$t^{\prime}$}
\put(107,33){\line(1,0){8}} 
\put(119,33){\line(1,0){8}}
\put(131,33){\line(1,0){8}}
\put(139,33){\circle*{3.5}}
\put(131,21){$t^{\prime\prime}$}
\put(139,33){\line(1,1){25}}
\put(139,33){\line(1,0){35}}
\put(139,33){\line(1,-1){25}}
\put(176,31){.}
\end{picture}
\end{center}
\vspace{-0.2cm}
The factor 3 (called {\it symmetry factor}) is due to the fact that the vertex in $t^{\prime\prime}$ can be attached to each one of the 3 continuous legs coming out from the vertex in $t^{\prime}$. It is interesting to note that in this diagrammatics the only  contribution to the solutions of the equations of motion is given by tree diagrams. This is due to the fact that the propagator $G_{\textrm{\scriptsize{R}}}(t-t^{\prime})$ is proportional to $\sin \omega(t-t^{\prime})$, so it cannot generate loops because it is identically zero for $t=t^{\prime}$.
These tree diagrams are similar to the Wyld diagrams introduced in the theory of turbulence, see for example \cite{mccomb}. 
 
A more compact proof of the equivalence between our perturbation theory via Feynman diagrams and the more standard secular perturbation theory goes as follows. 
It is easy to realize that the following iterative relation holds:
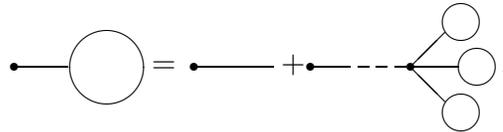
\begin{center}
\begin{picture}(150,60)
\put(20,30){\circle{28}}
\put(-15,30){\line(1,0){20}}
\put(-15,30){\circle*{3}}
\put(37,28){=}
\put(53,30){\line(1,0){30}}
\put(53,30){\circle*{3}}
\put(86,28){+}
\put(97,30){\line(1,0){15}}
\put(97,30){\circle*{3}}
\put(115,30){\line(1,0){4}}
\put(122,30){\line(1,0){4}}
\put(129,30){\line(1,0){4}}
\put(135,30){\circle*{3}}
\put(135,30){\line(1,1){13}}
\put(135,30){\line(1,0){18}}
\put(135,30){\line(1,-1){13}}
\put(160,30){\circle{14}}
\put(154,47){\circle{14}}
\put(154,13){\circle{14}}
\put(169,28){.}
\end{picture}
\end{center}
The equation associated with the previous diagram is:
\begin{displaymath}
\displaystyle q(t)=q_0(t)- \frac{g}{6}\int_{0}^{T}\textrm{d}t^{\prime}\, G_{\textrm{\scriptsize{R}}}(t-t^{\prime})q^3(t^{\prime}).
\end{displaymath}
By deriving the previous equation twice w.r.t. $t$ and using (\ref{B2bis}), we get exactly the equation of motion (\ref{eqmotion}) of the anharmonic oscillator. This completes the proof of the equivalence of our diagrammatics with the standard secular approach to classical perturbation theory.

\section{Conclusions}

In this paper we have shown how Feynman diagrams can be used also in classical mechanics as a useful stenographic tool to implement the standard secular perturbation theory. Of course, the formalism that we have presented here can be easily modified to take into account other examples. In particular, a change in the free generating functional implies a change in the form of the propagator, while a change in the perturbation potential $V(q)$ implies a change in the form of the vertex. 
We think that this approach to perturbation theory can be used to convey to students the idea that path integrals and Feynman diagrams are powerful tools, which can be fruitfully employed not only in quantum field theories but also in other areas. 

\appendix
\makeatletter
\@addtoreset{equation}{section}
\makeatother

\section{}
In this Appendix we want to prove explicitly Eq. (\ref{firstappendix}). Let us start from the definition of the generating functional for classical mechanics (\ref{zeta}):
\begin{displaymath}
\displaystyle Z[J,\Lambda]\equiv \int \textrm{d}\varphi_f {\mathcal D}^{\prime\prime}\varphi{\mathcal D}\lambda \, e^{i\int_{0}^{T}\textrm{\scriptsize{d}}t\, ({\mathcal L}+Jq+\Lambda\lambda_p)}.
\end{displaymath}
Calculating the functional derivative of $Z$ w.r.t. $J(t^{\prime})$, and putting afterwards  the currents $J$ and $\Lambda$ equal to zero, we get:
\begin{displaymath}
\displaystyle -i\frac{\delta Z[J,\Lambda]}{\delta J(t^{\prime})}\Biggl|_{J,\Lambda=0}=
\int {\textrm{d}}\varphi_f{\mathcal D}^{\prime\prime}\varphi{\mathcal D}\lambda \, q(t^{\prime}) \, e^{i\int_{0}^{T}\textrm{\scriptsize{d}}t\, {\mathcal L}}.
\end{displaymath}
As usual, the functional integral over $\lambda$ can be performed explicitly to get the functional Dirac delta of the equations of motion:
\begin{displaymath}
\displaystyle  -i\frac{\delta Z[J,\Lambda]}{\delta J(t^{\prime})}\Biggl|_{J,\Lambda=0}=
\int \textrm{d}\varphi_f {\mathcal D}^{\prime\prime}\varphi \, q(t^{\prime}) \, \delta(\dot{\varphi}^a-\omega^{ab}\partial_bH).
\end{displaymath}
Next, we can use the analog of Eq. (\ref{zeros}) and pass from the delta of the equations of motion to the one of the solutions of the equations $\delta[\varphi^a(t)-\varphi^a_{\textrm{\scriptsize{cl}}}(t;\varphi_i)]$. Also the path integral over $\varphi$ can be performed explicitly and we get:
\begin{eqnarray*}
\displaystyle -i\frac{\delta Z[J,\Lambda]}{\delta J(t^{\prime})}\Biggl|_{J,\Lambda=0}&=&
q_{\textrm{\scriptsize{cl}}}(t^{\prime};\varphi_i) \int \textrm{d}\varphi_f \delta[\varphi^a_f-\varphi^a_{\textrm{\scriptsize{cl}}}(T;\varphi_i)]\medskip \\
&=& q_{\textrm{\scriptsize{cl}}}(t^{\prime};\varphi_i).
\end{eqnarray*}
This completes the proof of Eq. (\ref{firstappendix}).

\section{}
In this Appendix we want to prove Eq. (\ref{zedzero}), i.e. we want to derive the generating functional $Z_0[J,\Lambda]$ for the harmonic oscillator. First of all, let us introduce a new Hamiltonian $H_0^{\prime}\equiv H_0+\Lambda q$. With this definition the generating functional of the harmonic oscillator can be written as follows:
\begin{eqnarray*}
\displaystyle Z_0[J,\Lambda]&=& \int \textrm{d}\varphi_f  {\mathcal D}^{\prime\prime}\varphi {\mathcal D}\lambda \,
e^{i \int_{0}^{T}\textrm{\scriptsize{d}}t \, (\lambda_a\dot{\varphi}^a-\lambda_a\omega^{ab}\partial_bH_0^{\prime}+Jq)}\nonumber \\
&=& \int \textrm{d}\varphi_f {\mathcal D}^{\prime\prime} \varphi \, \delta(\dot{\varphi}^a-\omega^{ab}\partial_bH_0^{\prime}) \, e^{i\int_{0}^{T}\textrm{\scriptsize{d}}t \, Jq},
\end{eqnarray*}
where in the last step we have performed the path integral over $\lambda$. 
Using the analog of Eq. (\ref{zeros}) for the Hamiltonian $H_0^{\prime}$ and disregarding the functional determinant, which is independent of the variables $q$ and $\lambda_p$, we can rewrite the generating functional as:
\begin{eqnarray}
\displaystyle Z_0[J,\Lambda]&=&{\mathcal N} \int \textrm{d}\varphi_f{\mathcal D}^{\prime\prime}\varphi \, \delta[\varphi^a-\varphi^{\prime a}_{\textrm{\scriptsize{cl}}}(t;\varphi_i)] \, e^{i\int_{0}^{T}\textrm{\scriptsize{d}}t \, Jq} \nonumber\\
&=& {\mathcal N} \int \textrm{d}\varphi_f \, \delta[\varphi_f-\varphi^{\prime}_{\textrm{\scriptsize{cl}}}(T;\varphi_i)] \, e^{i\int_{0}^{T}\textrm{\scriptsize{d}}t\, J q^{\prime}_{\textrm{\scriptsize{cl}}}} \nonumber  \\
&=& {\mathcal N}\, \exp \left( i\int_{0}^{T} \textrm{d}t\, Jq^{\prime}_{\textrm{\scriptsize{cl}}} \right). \label{zerojk} \\
\nonumber
\end{eqnarray}
The condition $Z[J,\Lambda=0]=1$ fixes ${\mathcal N}=1$. Finally, we have to find the explicit form of $q^{\prime}_{\textrm{\scriptsize{cl}}}$ by solving the equations of motion associated with $H_0^{\prime}$, i.e.: 
\begin{eqnarray*}
\dot{q}=p, \qquad
\dot{p}=-\omega^2 q-\Lambda,
\end{eqnarray*}
which are equivalent to the following second order equation for $q$:
\begin{equation}
\displaystyle \ddot{q}+\omega^2q=-\Lambda.  \label{inh}
\end{equation}
Let us consider the so called retarded Green function $G_{\textrm{\scriptsize{R}}}(t-t^{\prime})$ of Eq. (\ref{green}), which satisfies the following equation:
\begin{equation}
\displaystyle (\partial_t^2+\omega^2)G_{\textrm{\scriptsize{R}}}(t-t^{\prime})=\delta(t-t^{\prime}) \label{B2bis}
\end{equation}
with the boundary condition $G_{\textrm{\scriptsize{R}}}(t-t^{\prime})=0$ for $t<t^{\prime}$.
Then Eq. (\ref{inh}) can be solved in terms of the Green function as follows:
\begin{equation}
\displaystyle q^{\prime}_{\textrm{\scriptsize{cl}}}(t)=q_{0}(t)- \int_0^T \textrm{d}t^{\prime}\, G_{\textrm{\scriptsize{R}}}(t-t^{\prime}) \Lambda(t^{\prime}),  \label{zerojk2}
\end{equation}
where $q_{0}(t)$ is the solution (\ref{harm}) of the homogenous equation $\ddot{q}+\omega^2q=0$ with initial conditions $q(0)=q_i$ and $\dot{q}(0)=p_i$.
Finally, replacing (\ref{zerojk2}) into (\ref{zerojk}), we get the explicit form (\ref{zedzero}) of the generating functional for a harmonic oscillator. 

\section*{Acknowledgements}
We would like to thank E. Gozzi for having turned our attention to this topic and for many helpful suggestions.
This research has been supported by grants from INFN, MIUR and the University of Trieste.


\section*{References}
\begin{thebibliography}{99}
\bibitem{hibbs} R.P. Feynman and A.R. Hibbs, ``{\it Quantum mechanics and path integrals}" (Mc Graw-Hill, New York, 1965).
\bibitem{ennio} E. Gozzi, M. Reuter and W.D. Thacker, Phys. Rev. D {\bf 40} (1989) 3363. 
\bibitem{book}
A. Das, ``{\it Field theory: a path integral approach}" (World Scientific, Singapore, 1993).
\bibitem{jose'} J.V. Jos\'e and E.J. Saletan, ``{\it Classical dynamics: a contemporary approach}" (Cambridge University Press, 1998). 
\bibitem{mccomb}
W.D. McComb, ``{\it Renormalization methods: a guide for beginners}" (Oxford University Press, 2004). 
\end{thebibliography}
\end{document}